\documentclass[twocolumn]{aastex701}
\usepackage[utf8]{inputenc}

\usepackage{amsmath}
\usepackage{amssymb}
\usepackage{newtxtext}
\usepackage{newtxmath}
\usepackage[style=american]{csquotes}
\usepackage{comment}
\usepackage{physics}
\let\tablenum\relax
\usepackage{siunitx}
\newcommand{\numobs}{s}
\newcommand{\numtrial}{n}
\newcommand{\occrate}{\eta}

\newcommand{\uniformdist}{\mathcal{U}}
\newcommand{\kepler}{\emph{Kepler}}

\newcommand{\ktwo}{\emph{K2}}
\newcommand{\tess}{\emph{TESS}}
\newcommand{\gaia}{\emph{Gaia}}
\newcommand{\hj}{hot jupiter}
\newcommand{\planet}[2]{#1\thinspace #2}
\newcommand{\fiftyonepegb}{\planet{51\thinspace Peg}{b}}
\newcommand{\waspfortyseven}{WASP-47}
\newcommand{\waspfortysevenhj}{\planet{\waspfortyseven}{b}}
\newcommand{\waspfortyseveninner}{\planet{\waspfortyseven}{e}}
\newcommand{\waspfortysevenouter}{\planet{\waspfortyseven}{d}}
\newcommand{\waspfortysevencold}{\planet{\waspfortyseven}{c}}
\newcommand{\keplerseventhirty}{Kepler-730}
\newcommand{\keplerseventhirtyhj}{\planet{Kepler-730}{b}}
\newcommand{\keplerseventhirtyinner}{\planet{Kepler-730}{c}}
\newcommand{\toieleventhirty}{TOI-1130}
\newcommand{\toieleventhirtyinner}{\planet{\toieleventhirty}{b}}

\newcommand{\toitwothousand}{TOI-2000}
\newcommand{\toitwothousandinner}{\planet{\toitwothousand}{b}}

\newcommand{\wasponethirtytwo}{WASP-132}
\newcommand{\wasponethirtytwohj}{\planet{\wasponethirtytwo}{b}}
\newcommand{\wasponethirtytwoinner}{\planet{\wasponethirtytwo}{c}}
\newcommand{\waspeightyfour}{WASP-84}

\newcommand{\waspeightyfourinner}{\planet{\waspeightyfour}{c}}
\newcommand{\toififtyonefortythree}{TOI-5143}

\newcommand{\toififtyonefortythreeinner}{\planet{\toififtyonefortythree}{b}}
\newcommand{\toitwentyfourninetyfour}{TOI-2494}

\newcommand{\toitwentyfourninetyfourinner}{\planet{\toitwentyfourninetyfour}{b}}
\newcommand{\toifourteenoheight}{TOI-1408}

\newcommand{\toifourteenoheightinner}{\planet{\toifourteenoheight}{c}}
\newcommand{\toifortyfoursixtyeight}{TOI-4468}
\newcommand{\toifortyfoursixtyeighthj}{\planet{\toifortyfoursixtyeight}{b}}
\newcommand{\toifortyfoursixtyeightouter}{\planet{\toifortyfoursixtyeight}{c}}
\DeclareSIUnit{\mass}{M}
\DeclareSIUnit{\radius}{R}
\DeclareSIUnit{\year}{yr}
\DeclareSIQualifier{\sun}{\ensuremath{\odot}}
\DeclareSIQualifier{\earth}{\ensuremath{\oplus}}
\DeclareSIQualifier{\jupiter}{J}
\DeclareSIPostPower{\nominal}{N}
\newcommand{\bjdtdb}{\ensuremath{\mathrm{BJD}_\text{TDB}}}

\newcommand{\snr}{\ensuremath{\mathrm{S}/\mathrm{N}}}
\newcommand{\transitdur}{T_\text{dur}}
\newcommand{\detecteff}{\varepsilon}
\newcommand{\astropy}{\textsc{Astropy}}
\newcommand{\batman}{\textsc{Batman}}
\newcommand{\cuvarbase}{\textsc{cuvarbase}}
\newcommand{\giants}{\textsc{giants}}

\newcommand{\vartools}{\textsc{Vartools}}

\newcommand{\occRateHjAll}{$(7.6^{+5.5}_{-3.8})\%$}
\newcommand{\occRateHjIsotropic}{$(37^{+19}_{-17})\%$}
\newcommand{\occRateHjAligned}{$(7.4^{+5.4}_{-3.7})\%$}

\newcommand{\ut}{\textsc{ut}}

\shorttitle{The Occurrence Rate of Nearby Planetary Companions to Hot Jupiters}
\shortauthors{Sha et al.}

\begin{document}

\title{The Occurrence Rate of Nearby Planetary Companions to Hot Jupiters}
\received{October 3, 2025}
\revised{June 15, 2026}
\accepted{June 18, 2026}
\submitjournal{AJ}

\suppressAffiliations
\author[0000-0001-5401-8079]{Lizhou Sha}
\affiliation{Department of Astrophysical Sciences, Princeton University, 4 Ivy Ln, Princeton, NJ 08540 USA}
\email{user@example.edu}

\author[0000-0001-7246-5438]{Andrew M. Vanderburg}
\affiliation{Center for Astrophysics \textbar \ Harvard \& Smithsonian, 60 Garden Street, Cambridge, MA 02138, USA}
\email{user@example.edu}

\author[0000-0003-0918-7484]{Chelsea X. Huang}
\altaffiliation{ARC Future Fellow}
\affiliation{Centre for Astrophysics, University of Southern Queensland, Toowoomba, Queensland 4350, Australia}
\email{user@example.edu}

\author[0000-0003-0046-2494]{Samuel Christian}
\affiliation{Department of Physics and Kavli Institute for Astrophysics and Space Research, Massachusetts Institute of Technology, 77 Massachusetts Ave, Cambridge, MA 02139, USA}
\email{user@example.edu}

\author[0000-0002-7226-836X]{Kevin Burdge}
\affiliation{Department of Physics and Kavli Institute for Astrophysics and Space Research, Massachusetts Institute of Technology, 77 Massachusetts Ave, Cambridge, MA 02139, USA}
\email{kburdge@mit.edu}

\author[0000-0003-2657-3889]{Nicholas Saunders}
\affiliation{Department of Astronomy, Yale University, New Haven, CT 06511, USA}
\affiliation{Institute for Astronomy, University of Hawaii at M\=anoa, 2680 Woodlawn Drive, Honolulu, HI 96822, USA}
\email{saunders.nk@gmail.com}

\author[0000-0003-1464-9276]{Khalid Barkaoui}
\affiliation{Instituto de Astrofísica de Canarias (IAC), E-38200 La Laguna, Tenerife, Spain}
\affiliation{Astrobiology Research Unit, Universit\'e de Li\`ege, All\'ee du 6 Ao\^ut 19C, B-4000 Li\`ege, Belgium}
\affiliation{Department of Earth, Atmospheric and Planetary Science, Massachusetts Institute of Technology, 77 Massachusetts Ave, Cambridge, MA 02139, USA}
\email{user@example.edu}

\author[0000-0003-3469-0989]{Alexander Belinski}
\affiliation{Sternberg Astronomical Institute, Lomonosov Moscow State University, Universitetskii prospekt, 13, Moscow 119992, Russia}
\email{user@example.edu}

\author[0000-0003-0846-1744]{Serge Bergeron}
\affiliation{American Public University}
\affiliation{American Association of Variable Star Observers, 185 Alewife Brook Pkwy Ste 410, Cambridge, MA 02138, USA}
\email{astrosberge@gmail.com}

\author[0000-0002-7564-6047]{Zo{\"e}\ L. de Beurs}
\affiliation{Department of Earth, Atmospheric and Planetary Sciences, Massachusetts Institute of Technology, Cambridge, MA 02139, USA}
\altaffiliation{NSF Graduate Research Fellow, MIT Presidential Fellow, MIT Collamore-Rogers Fellow}
\email{zdebeurs@mit.edu}

\author[0000-0001-6637-5401]{Allyson Bieryla}
\affiliation{Center for Astrophysics \textbar \ Harvard \& Smithsonian, 60 Garden Street, Cambridge, MA 02138, USA}
\email{user@example.edu}

\author[0000-0001-6588-9574]{Karen A. Collins} 
\affiliation{Center for Astrophysics \textbar \ Harvard \& Smithsonian, 60 Garden Street, Cambridge, MA 02138, USA}
\email{user@example.edu}

\author[0000-0002-2412-1558]{Giuseppe Conzo}
\affiliation{Gruppo Astrofili Palidoro, Italy}
\email{user@example.edu}

\author[0000-0003-0597-7809]{Gareb Fern\'andez-Rodr\'iguez}
\affiliation{Instituto de Astrofísica de Canarias (IAC), E-38200 La Laguna, Tenerife, Spain}
\affiliation{Departamento de Astrofísica, Universidad de La Laguna (ULL), E-38206 La Laguna, Tenerife, Spain}
\email{gareb.fernandez@gmail.com}

\author[0000-0002-4909-5763]{Akihiko Fukui}
\affiliation{Komaba Institute for Science, The University of Tokyo, 3-8-1 Komaba, Meguro, Tokyo 153-8902, Japan}
\affiliation{Instituto de Astrofísica de Canarias (IAC), E-38200 La Laguna, Tenerife, Spain}
\email{afukui@g.ecc.u-tokyo.ac.jp}

\author[0000-0001-8627-9628]{Davide Gandolfi}
\affiliation{Dipartimento di Fisica, Università degli Studi di Torino, via Pietro Giuria 1, 10125, Torino, Italy}
\email{davide.gandolfi@unito.it}

\author[0000-0002-7188-8428]{Tristan Guillot}
\affiliation{Observatoire de la C\^ote d'Azur, Universit\'e C\^ote d’Azur, CNRS, Laboratoire Lagrange, Bd de l'Observatoire, CS 34229, 06304 Nice cedex 4, France}
\email{tristan.guillot@oca.eu}

\author[0000-0001-8732-6166]{Joel D. Hartman}
\affiliation{Department of Astrophysical Sciences, Princeton University, 4 Ivy Ln, Princeton, NJ 08540 USA}
\email{jhartman@astro.princeton.edu}

\author[0000-0002-5978-057X]{Kai Ikuta}
\affiliation{Department of Social Data Science, Hitotsubashi University, 2-1 Naka, Kunitachi, Tokyo 186-8601, Japan}
\email{user@example.edu}

\author[0000-0001-9911-7388]{David W. Latham} 
\affiliation{Center for Astrophysics \textbar \ Harvard \& Smithsonian, 60 Garden Street, Cambridge, MA 02138, USA}
\email{user@example.edu}

\author[0000-0002-6424-3410]{Jerome P. de Leon}
\affiliation{Komaba Institute for Science, The University of Tokyo, 3-8-1 Komaba, Meguro, Tokyo 153-8902, Japan}
\email{user@example.edu}

\author[0000-0001-8879-7138]{Bob Massey}
\affiliation{American Association of Variable Star Observers, 185 Alewife Brook Pkwy Ste 410, Cambridge, MA 02138, USA}
\email{user@example.edu}

\author[0000-0001-7809-1457]{Gabriel Murawski}
\affiliation{Gabriel Murawski Private Observatory (SOTES)}
\email{user@example.edu}

\author[0000-0001-9087-1245]{Felipe Murgas}
\affiliation{Instituto de Astrofísica de Canarias (IAC), E-38200 La Laguna, Tenerife, Spain}
\affiliation{Departamento de Astrofísica, Universidad de La Laguna (ULL), E-38206 La Laguna, Tenerife, Spain}
\email{user@example.edu}

\author[0000-0001-8511-2981]{Norio Narita}
\affiliation{Komaba Institute for Science, The University of Tokyo, 3-8-1 Komaba, Meguro, Tokyo 153-8902, Japan}
\affiliation{Astrobiology Center, 2-21-1 Osawa, Mitaka, Tokyo 181-8588, Japan}
\affiliation{Instituto de Astrofísica de Canarias (IAC), E-38200 La Laguna, Tenerife, Spain}
\email{user@example.edu}

\author[0000-0002-8986-6681]{Mohammad Odeh}
\affiliation{International Astronomical Center, 45015, Abu Dhabi, United Arab Emirates}
\email{user@example.edu}

\author[0000-0003-0987-1593]{Enric Palle}
\affiliation{Instituto de Astrofísica de Canarias (IAC), E-38200 La Laguna, Tenerife, Spain}
\affiliation{Departamento de Astrofísica, Universidad de La Laguna (ULL), E-38206 La Laguna, Tenerife, Spain}
\email{user@example.edu}

\author[0000-0001-5519-1391]{Hannu Parviainen}
\altaffiliation{Ram\'on y Cajal Fellow}
\affiliation{Instituto de Astrofísica de Canarias (IAC), E-38200 La Laguna, Tenerife, Spain}
\affiliation{Departamento de Astrofísica, Universidad de La Laguna (ULL), E-38206 La Laguna, Tenerife, Spain}
\email{hannu@iac.es}

\author[0009-0006-7023-1199]{Gabrielle Ross}
\affiliation{Department of Astrophysical Sciences, Princeton University, 4 Ivy Ln, Princeton, NJ 08540 USA}
\email{gr8740@princeton.edu}

\author[0000-0001-8227-1020]{Richard P. Schwarz}
\affiliation{Center for Astrophysics \textbar \ Harvard \& Smithsonian, 60 Garden Street, Cambridge, MA 02138, USA}
\email{user@example.edu}

\author{Gregor Srdoc}
\affiliation{Kotizarovci Observatory, Sarsoni 90, 51216 Viskovo, Croatia}
\email{user@example.edu}

\author[0000-0003-2163-1437]{Chris Stockdale}
\affiliation{Hazelwood Observatory, Australia}
\email{thestockdalefamily@bigpond.com}

\author[0000-0002-8964-8377]{Samuel N. Quinn}
\affiliation{Center for Astrophysics \textbar \ Harvard \& Smithsonian, 60 Garden Street, Cambridge, MA 02138, USA}
\email{squinn@cfa.harvard.edu}

\author[0000-0002-3249-3538]{Ian A. Waite} 
\affiliation{Centre for Astrophysics, University of Southern Queensland, Toowoomba, Queensland 4350, Australia}
\email{user@example.edu}

\author[0000-0003-2127-8952]{Francis P. Wilkin}
\affiliation{Union College, Schenectady, NY 12308 USA}
\email{user@example.edu}

\author[0000-0002-4891-3517]{George Zhou}
\altaffiliation{ARC Future Fellow}
\affiliation{Centre for Astrophysics, University of Southern Queensland, Toowoomba, Queensland 4350, Australia}
\email{george.zhou@unisq.edu.au}

\begin{abstract}
Of the $> 500$ confirmed transiting hot jupiters
and $\approx 2000$ additional candidates today,
only ten are known to have nearby companion planets.
The survival of nearby companions means that these hot jupiters cannot have migrated to their present location via dynamically disruptive high-eccentricity migration
but instead have undergone disk migration or formed in situ.
The occurrence rate for these nearby companions, therefore, constrains the relative efficiency of different hot jupiter formation pathways.
Here, we perform a uniform box least-squares search for nearby transiting companions to hot jupiters in the first five years of TESS data.
Accounting for observational completeness and detection efficiency, 
we arrive at an occurrence rate of
\occRateHjAll{},
which is a lower limit on the fraction of hot jupiters that underwent disk migration or in situ formation.
Comparing this rate with that derived from transit-timing variation searches
suggests that hot jupiters are likely mostly aligned with their nearby companions,
but their apparently higher incidence of grazing transits
may point to a slight preferential misalignment.
We also synthesize evidence that hot jupiters with nearby companions
may have cold companions at a rate similar to that of other hot jupiters.
Comprehensive transit, radial velocity, and stellar obliquity measurements
in hot jupiter systems with nearby companions will be necessary
to fully account for the relative prevalence of proposed hot jupiter formation pathways.
\end{abstract}

\section{Introduction}

The existence of nearby planetary companions to hot jupiters in the present
is a powerful diagnostic of their formation in the past.
Although hot jupiters,
or gas giant planets with orbital periods less than 10~days,
were among the first exoplanets ever discovered,
we have yet to complete the theoretical accounting of their formation.
Under the core accretion paradigm \citep[e.g.][]{poll96},
gas giant planets like Jupiter and Saturn should only form at large orbital separations from their host stars,
beyond \enquote{snow lines} where volatile materials exist in solid form
within the protoplanetary disk.
The discovery of the prototypical hot jupiter \fiftyonepegb{}
\citep{Mayor1995Natur}, therefore, initially came as a surprise.
While an alternative is for hot jupiters to form in situ near where they are found today
after being seeded by super-earth-sized planets \citep{Batygin2016ApJ},
a majority of the literature resolves this apparent conflict
by proposing that they formed \enquote{cold} via core accretion and then migrated to their present \enquote{hot} location.
Whether nearby planetary companions to hot jupiters can exist
depends on how this migration unfolds.

Two main migration pathways have been proposed for hot jupiters:
high-eccentricity (or tidal) migration \citep[e.g.][]{Rasio1996Sci} and disk migration \citep[e.g.][]{Lin1996Natur}. 
Under high-eccentricity migration, a Jupiter-sized planet that formed beyond the snow line is perturbed into a highly eccentric orbit by gravitational interactions with an outer planetary or stellar companion.
From there, tides raised on the planet during periastron passage gradually dissipate its orbital energy and shrink its semimajor axis
until its orbit circularizes.
Under disk migration, on the other hand,
the hot jupiter forms before the protoplanetary disk completely dissipates
and interacts with the disk to gradually reduce its orbital period.
Understanding in what proportion these two migration pathways (along with in situ formation)
are responsible for the known population of hot jupiters
is a major goal of exoplanet science. 

One way to discriminate these formation pathways is to look for smaller planets orbiting interior to hot jupiters.
Because high-eccentricity migration is dynamically disruptive, it is expected that no inner companion could have survived the process
\citep{2015ApJ...808...14M}.
If an inner companion is found, we could all but eliminate high-eccentricity migration as a mechanism by which that hot jupiter could have formed.
Thus, studying the population of nearby planetary companions to hot jupiters presents an opportunity to constrain the fraction of these systems that formed via dynamically quiet pathways.

Early estimates of the occurrence rate of nearby planetary companions to hot jupiters
were hampered by a limited sample of transiting hot jupiters
observed with sufficient photometric precision to detect small companion planets.
For 20 years after the discovery of \fiftyonepegb{}, no hot jupiters were found to have nearby planetary companions \citep[e.g.][]{Miller-Ricci2008ApJ, Steffen2012PNAS},
even while smaller planets were frequently found in multiplanet systems \citep{lath11}
by the \kepler{} prime mission.
Subsequently during the \ktwo{} mission, the discovery of nearby planetary companions to the hot jupiter \waspfortysevenhj{} \citep{2015ApJ...812L..18B} showed that at least some hot jupiters form in a dynamically quiet manner, either via disk migration or \textit{in situ}.
This discovery prompted the first estimate of the occurrence rate
of nearby companions to hot jupiters
($1.1^{+13.3}_{-1.1}\%$, median and 90\% interval)
by \citet*{2016ApJ...825...98H},
suggesting that those systems were rare,
although this rate was based on null detection during the \kepler{} prime mission at the time.
Similarly, after the discovery of \keplerseventhirtyinner{} by reanalysis of \kepler{} prime mission planet candidates
\citep{Zhu2018RNAAS,Canas2019ApJL},
\citet{2021ARA&A..59..291Z} also made an independent estimate based on the final \kepler{} DR25 planet candidate catalog \citep{2018ApJS..235...38T},
counting \keplerseventhirty{} as a detection,
and arrived at a similar occurrence rate ($\approx 2\%$, or $<9.7\%$ at 95\% upper limit)
for nearby ($P \lesssim \SI{20}{\day}$), small ($\approx$1--\SI{4}{\radius\earth}), and nearly coplanar companions to hot jupiters.

The apparent loneliness of hot jupiters suggests that high-eccentricity migration is likely the dominant formation pathway.
Nevertheless, the relatively large uncertainty on the occurrence rate, a consequence of the relatively small number of hot jupiters searched, limits the strength of conclusions about hot jupiter formation that can be drawn. 
Thanks to its wide field of view, nearly full-sky coverage, and full-frame images with sufficient cadence,
NASA's Transiting Exoplanet Survey Satellite (\tess{})
has brought a flood of hot jupiter candidates.
Hot jupiters are intrinsically rare \citep{Wright2012ApJ}, and it is difficult to construct large statistical samples without this comprehensive sky coverage and high data volume.
Before \tess{}, the best statistical sample of hot jupiters was from the \kepler{} mission, but because \kepler{} was only able to downlink data for about 200,000 stars, it only detected 40--60 hot jupiters from its primary mission \citep{2018ApJS..235...38T,Yee2021AJ}.
\tess{}, on the other hand, observes and downlinks data for millions of stars bright enough for hot jupiter detection each month.
As a result, it has already observed thousands of \hj s \citep{2021ApJS..254...39G},
the vast majority of which are new discoveries \citep[e.g.][]{Yee2022AJ, Yee2023ApJS, Rodriguez2021AJ, Rodriguez2023MNRAS, Schulte2024AJ}.
This large number of discoveries, coupled with TESS's very high sensitivity to these planets (making large magnitude- and volume-limited samples nearly complete) and relatively unbiased target selection, has enabled important advancements in hot jupiter population studies
\citep[e.g.][]{2019AJ....158..141Z, Beleznay2022MNRAS, Yee2023ApJL}.
Among these results was the measurement of the occurrence rate of nearby companions around hot jupiters by \citet{Hord2021AJ}, who used a sample of about 179 hot jupiters to determine that about $7.3^{+15.2}_{-7.3}$\% (median and 90\% interval, or $< 27.8\%$ at 95th-percentile upper limit) of hot jupiters have a nearby planetary companion.

\begin{figure}
    \centering
    \includegraphics[width=\columnwidth]{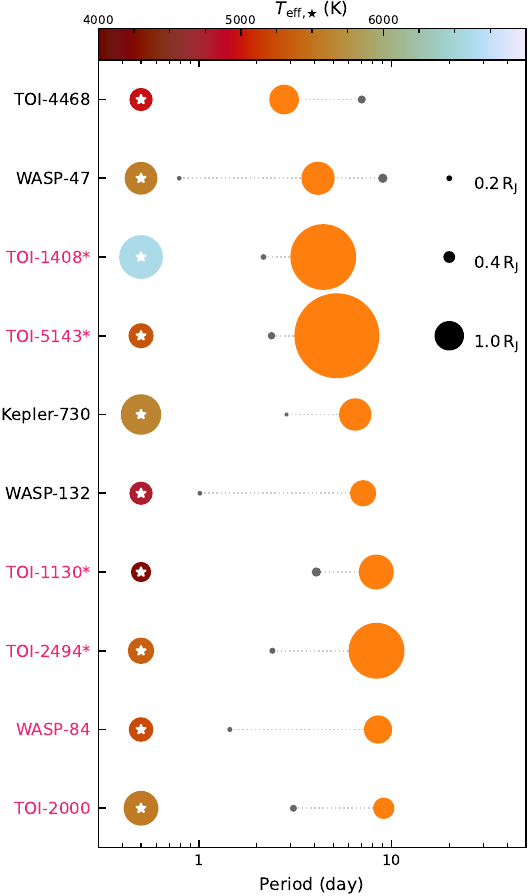}
    \caption{\added{Transiting planetary systems hosting nearby planetary companions to hot jupiters ($P < 10\,\text{d}$).
    Non-transiting planets are omitted.
    The leftmost circle in each row represents the host star,
    with the mark's size indicating the stellar radius
    and fill color the stellar effective temperature
    \citep[from TICv8;][]{2019AJ....158..138S}.
    The other circles represent the transiting planets in each system,
    with their sizes proportional to planet radii
    and fill colors indicating a giant planet (orange) or a small planet (gray).
    The relative mark sizes among either the stars or the planets are to scale,
    but not between a star and its planets.
    The translucent marks are planet candidates not yet published in the literature.
    Systems which are counted as detected (\autoref{sec:results}) for the occurrence rate calculation (\autoref{sec:occurrence}) have their names highlighted in magenta,
    and systems whose giant planets have grazing transits (\autoref{sec:grazing}) are marked with an asterisk.
    The systems are sorted in ascending order of the period of the largest planet.
    This figure and its caption are adapted from Figure~10 by \citet{2023MNRAS.524.1113S},
    updated with newly available planetary parameters
    for the \toifortyfoursixtyeight{} \citep{Livesey2026ApJL},
    \toitwentyfourninetyfour{}, \toififtyonefortythree{} \citep{2026AJ....171..359Q},
    and \toifourteenoheight{} \citep{2024ApJ...971L..28K} systems.}}
    \label{fig:family_portrait}
\end{figure}

Our work builds on the large sample of hot jupiters observed by \tess{}
and previous demographic studies of nearby companions to hot jupiters.
Since the work by
\citet{Hord2021AJ}, several other hot jupiter systems with nearby companions have been discovered
\citep{2020ApJ...892L...7H, 2022AJ....164...13H, 2023MNRAS.525L..43M, 2023MNRAS.524.1113S, 2024ApJ...971L..28K, McKee2025ApJ, 2026AJ....171..359Q, Livesey2026ApJL}.
This means that, for the first time, there is an opportunity to measure the occurrence rate of these nearby companions
based on positive detections
and a large statistical sample of hot jupiters.
\autoref{fig:family_portrait} provides a \enquote{family portrait} of the architectures of these compact multiplanet hot jupiter systems.

In this work, we perform the largest ever uniform search for short-period transiting companions to transiting hot jupiters
and calculate the occurrence rate of these companions corrected for observational biases and detection completeness,
for the first time based on a positive number of detections.
We identify $\approx 2000$ likely hot jupiters from the \tess{} Objects of Interest catalog of planet candidates (\autoref{sec:sample}).
After searching their light curves for transiting companions with small orbital periods
(\autoref{sec:search}),
we carefully vet any new signal and eliminate known astrophysical false positives such as eclipsing binaries or stellar variations, reporting our detections in \autoref{sec:results}.
We characterize the sensitivity of our pipeline and calculate the occurrence rate for nearby companions, given the existence of these hot jupiters (\autoref{sec:occurrence}).
We discuss the implications of the occurrence rate measurements and architectures of these planetary systems in \autoref{sec:discussion}, concluding with \autoref{sec:conclusions}.

\section{Sample of Hot Jupiters}  \label{sec:sample}

We aimed to search every hot jupiter
observed during the first five years
(sectors~1--69)
of the \tess{} mission
for nearby planetary companions
($P < \SI{10}{\day}$).
To that end, we assembled a list of confirmed and candidate hot jupiters
from the \tess{} Objects of Interest \citep[TOI;][]{2021ApJS..254...39G} Catalog.
After carefully removing known or suspected non-planets,
we estimated the false positive rate (FPR) of the remaining planet candidates on our list.
Finally, we derived new stellar parameters from broadband photometry
as the basis for refined planetary parameters of the hot jupiters.
These procedures are detailed in the rest of this section.

\subsection{Selecting hot jupiters}  \label{sec:select}

\begin{deluxetable*}{rDDDDDDDDD}
    \tablecaption{Hot jupiter candidates from the \tess{} Objects of Interest (TOI) catalog.
    \label{tab:hj}}
    \tabletypesize{\footnotesize}
    \tablehead{%
        \colhead{TIC ID} &
        \multicolumn2c{Period} &
        \multicolumn2c{Mid-transit time} &
        \multicolumn2c{Transit depth} &
        \multicolumn2c{Duration} &
        \multicolumn2c{Stellar mass} &
        \multicolumn2c{Stellar radius} &
        \multicolumn2c{Planet radius} &
        \multicolumn2c{$a/R_{\star}$} &
        \multicolumn2c{Impact param.}
        \\
        &
        \multicolumn2c{d} &
        \multicolumn2c{$\bjdtdb - \num{2457000}$} &
        \multicolumn2c{} &
        \multicolumn2c{h} &
        \multicolumn2c{\si{\mass\sun}} &
        \multicolumn2c{\si{\radius\sun}} &
        \multicolumn2c{\si{\radius\earth}} &
        &
    }
    \decimals
    \startdata
4711 & 2.33481520 & 2358.079233 & 0.005310 & 1.86 & 1.202 & 1.432 & 11.4 & 5.499 & 0.912 \\
56884 & 3.91748200 & 3088.668722 & 0.008397 & 2.62 & 1.322 & 1.574 & 15.7 & 7.292 & 0.890 \\
160023 & 6.75655190 & 2358.443291 & 0.005200 & 3.44 & 1.214 & 1.622 & 12.8 & 9.893 & 0.847 \\
857186 & 3.81236140 & 2465.891746 & 0.018330 & 2.56 & 0.826 & 0.788 & 11.6 & 12.227 & 0.365 \\
1003831 & 1.65114558 & 1518.203428 & 0.003206 & 0.95 & 1.008 & 1.077 & 6.7 & 5.472 & 0.976 \\
1129033 & 1.36002828 & 2168.520849 & 0.019428 & 2.17 & 0.982 & 1.045 & 15.9 & 4.913 & 0.526 \\
    \ldots & \multicolumn2c{\ldots} & \multicolumn2c{\ldots} & \multicolumn2c{\ldots} & \multicolumn2c{\ldots} & \multicolumn2c{\ldots} & \multicolumn2c{\ldots} & \multicolumn2c{\ldots} & \multicolumn2c{\ldots} & \multicolumn2c{\ldots}
    \enddata
    \tablecomments{\added{%
    The first five columns are taken from the TOI Catalog
    \cite[][accessed on 2024 July 5]{2021ApJS..254...39G}.
    The stellar masses and radii are from this work,
    and the last three columns are imputed from the
    values in the preceding columns
    assuming circular orbits
    (\autoref{sec:params}).
    These parameters are used in \autoref{sec:occurrence}
    to calculate observational completeness.
    While the values in this table are self-consistent
    and valid in the statistical aggregate,
    they should not be used in work demanding high accuracy
    in the context of individual planetary systems.
    (This table is available in its entirety in machine-readable form in the online article.).}}
\end{deluxetable*}

To start,
we select TOI candidates 
\begin{enumerate}
    \item with valid \gaia{} DR2 cross-matches,
    \item periods $P < \SI{10}{\day}$,
    \item radii $8 \leq R_\text{p} / \si{\radius\earth} < 30$ within uncertainty
        \emph{and} measured to better than 50\%, and
    \item alerted on or before \ut{} 2024 May 30
\end{enumerate}
from the Exo\-FOP website \citep[accessed on 2024 July 5]{exofop5}.
To identify TOI candidates that are known or suspected to not be planets,
we eliminated targets with
\begin{enumerate}
    \item unfavorable dispositions of the \tess{} Follow Up Program working group (TFOPWG)%
        \footnote{Specifically, we eliminated planets designated as false positive (FP),
        false alert (FA),
        and ambiguous planetary candidates (APC) by the TFOPWG.}
        from the ExoFOP website (accessed on 2025 March 11),
    \item excess radial velocity (RV) scatter from \gaia{} DR3, and
    \item $R_\text{p} \ge \SI{30}{\radius\earth}$ and
        semimajor axis--to--stellar radius ratio $a / R_\star < 1$
        based on the parameters derived in \autoref{sec:params}.
\end{enumerate}
The TFOPWG dispositions are based on follow-up photometry, spectroscopy,
and high-resolution imaging
and identify TOI candidates that are known or suspected to be eclipsing binaries or other astrophysical false positives.
The RV scatters from \gaia{} DR3 eliminate suspected spectroscopic binaries.
We estimate the RV scatter $\sigma_\text{RV}$
by multiplying the reported RV uncertainty (\texttt{radial\_velocity\_error})
and the square root of the number of RV observations (\texttt{rv\_nb\_transits}).
Then, we use the following empirical formula
based on \gaia{} DR3 $G$ magnitude
to identify the targets with excess $\sigma_\text{RV}$:
\begin{equation}
    \frac{\sigma_\text{RV}}{\si{\kilo\meter\per\second}} >
    \begin{cases}
    2  & G < 10 ,\\
    2 \cdot 10^{0.4 (G - 10)}  & G \geq 10 .
    \end{cases}
\end{equation}
The TOI candidates that survive these checks comprise our sample of hot jupiters
searched for nearby companions,
as long as their \tess{} light curve is available.
They are tabulated in \autoref{tab:hj}.

\subsection{Estimation of false positive rate} \label{sec:fpr}

In order to ensure the accuracy of our calculated occurrence rate (\autoref{sec:occ}),
we want to estimate the FPR, defined as the fraction of non-planets, in our hot jupiter sample.
Starting from the original estimate for the FPR of TOIs from
\citet{2019AJ....158..141Z}
based on 10 detected false positives among 31 TOIs,
which we denote $\mathrm{FPR}_\text{Zhou} = 10 / 31 \approx 0.32$,
we expect to find $\mathrm{FPR}_\text{Zhou} \cdot N_\text{TOI}$
false positives
among the $N_\text{TOI} = 2684$ TOIs that match our initial period, radius, and alert time criteria.
We subsequently eliminated
570 confirmed or suspected non-planets based on TFOPWG dispositions,
249 suspected spectroscopic binaries based on excess $\sigma_\text{RV}$,
and \added{16} suspected non-planets with implausible $R_\text{p}$ or $a/R_\star$
for a combined total of $N_\text{FP} = \added{749}$.
In theory, this elimination leaves us with
\begin{equation}
    \mathrm{FPR}_\text{HJ} = \frac{\mathrm{FPR}_\text{Zhou} \cdot N_\text{TOI} - N_\text{FP}}{N_\text{TOI} - N_\text{FP}}
    \approx 0.060
\end{equation}
among the remaining
$1935$ hot jupiters in our sample.
This $\mathrm{FPR}_\text{HJ}$
will be used in \autoref{sec:occ} to adjust the final effective sample size of the hot jupiters
for the occurrence rate calculation.

\subsection{Properties of the hot jupiters and their host stars} \label{sec:params}

\begin{figure*}
    \centering
    \includegraphics[width=\textwidth]{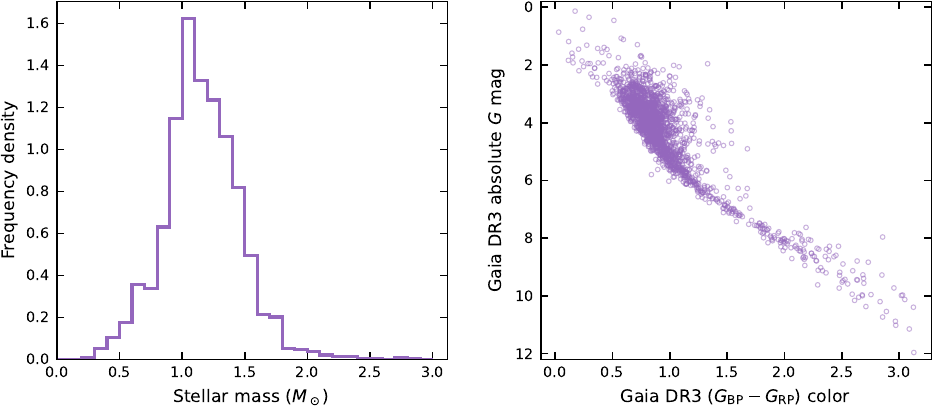}
    \caption{The distribution of the host stars of the selected hot jupiters.
    \emph{Left:} The histogram of stellar mass.
    \emph{Right:} The color--magnitude diagram.
    }
    \label{fig:star_dist}
\end{figure*}

In order to obtain more accurate planetary parameters,
we derived masses and radii for the hot jupiter host stars
via spectral energy distribution (SED) fitting
to \gaia{} DR3 $G$, $G_\text{BP}$, $G_\text{RP}$ magnitudes
and 2MASS $J$, $H$, $K$ magnitudes.
We obtained these magnitudes by cross-matching
the \tess{} Input Catalog version 8
\citep[TICv8;][]{2018AJ....156..102S,2019AJ....158..138S}
with the \gaia{} and 2MASS catalogs.
(The targets that did not have \gaia{} DR2 cross-matches in TICv8
were not selected in the first place.)
Given these broadband apparent magnitudes
and \gaia{} DR3 parallaxes,
we used the \textsc{isochrones} Python package \citep{morton2015}
to perform nested sampling of fundamental stellar parameters
modeled by the MESA Isochrones and Stellar Tracks
\citep[MIST;][]{2016ApJS..222....8D,2016ApJ...823..102C},
using the associated correction tables to convert bolometric to broadband photometric magnitudes.
To better reflect systematic and model uncertainty,
we inflated the uncertainty for all measured magnitudes to $0.1$.

The resulting new stellar masses $M_\star$ and radii $R_\star$,
as well as derived planetary parameters,
are tabulated in \autoref{tab:hj},
and the distribution of stellar masses and the color--magnitude diagram
are shown in \autoref{fig:star_dist}.
We derived planet radii $R_\text{p}$ from the new $R_\star$ and the transit depth from the TOI catalog
and used the new $M_\star$ and $R_\star$ combined with the orbital period from the TOI catalog
to calculate $a / R_\star$,
which is then used in conjunction with our derived $R_\text{p}$
and the transit duration from the TOI catalog
to calculate the impact parameter $b$, assuming circular orbits.
As mentioned in \autoref{sec:select},
we further identified all planets with
$R_\text{p} \ge \SI{30}{\radius\earth}$
and $a / R_\star < 1$
and eliminated them from the hot jupiter sample as likely non-planets.
Since TOIs must be transiting,
we clamped this derived $b$ to the interval $[0, 1 + R_\text{p} / R_\star)$.
These new $M_\star$, $R_\star$ and derived $R_\text{p}$, $b$ will be used
to determine observational completeness for the occurrence rate calculation in
\autoref{sec:occurrence}.

\section{Search for Additional Short-Period Planets in Hot Jupiter Systems} \label{sec:search}

To search for nearby transiting companions of hot jupiters,
we used light curves from the MIT Quick Look Pipeline (QLP)
up to and including sector~69.
Initially created for the \tess{} Prime Mission,
QLP generates light curves for all targets in the \tess{} full-frame images
(calibrated as described by \citealt{2020RNAAS...4..251F})
brighter than a \tess{} magnitude of 13.5
and certain fainter targets with high proper motion
\citep{2020RNAAS...4..204H,2020RNAAS...4..206H,2021RNAAS...5..234K,2022RNAAS...6..236K}.
Since we perform detrending, or removal of the long-term variability in the light curves,
as part of the planet search,
we use the raw simple aperture photometry (\texttt{SAP\_FLUX}) time series
for light curves before sector~56
and the systematics-corrected (\texttt{SYS\_RM\_FLUX}) time series thereafter
\citep{2022RNAAS...6..235K}.
There were $86$ targets out of the
$1935$ hot jupiter systems in our sample without valid QLP light curves,
leaving us with
$N_\text{HJ} = 1849$ systems to be searched.
We detail our planet search pipeline and candidate vetting procedures in this section.

\subsection{Box least-squares search}  \label{sec:bls}

We used the box least-squares
\citep*[BLS;][]{2002A&A...391..369K} algorithm
to search for additional planets in the light curves of the hot jupiters.
After removing all points that have been flagged for quality issues
(e.g. spacecraft anomaly, excess scattered light, etc.),
we iteratively applied the following steps five times to each light curve:
\begin{enumerate}
    \item Remove parts of the light curve containing hot jupiter transits and signals corresponding to the previous peaks of the light curve.
    \item Detrend the light curve, remove upward outliers, and estimate uncertainty.
    \item Find the period and phase of the signal corresponding to the peak of the BLS power spectrum.
\end{enumerate}
We detail these steps below.

\paragraph{Removal of known signals}
We remove the hot jupiter transits from the raw photometry light curve
by masking points that lie within a span of three times the transit duration centered on the mid-transit time,
with the transit ephemerides taken from the TOI Catalog.
In subsequent iterations, we also remove portions of the light curve
corresponding to signals detected in previous iterations.

\paragraph{Detrending}
We detrend the light curves using Keplerspline, a high-pass filtering method based on B-spline
\citep{vanderburg_technique_2014, shallue_identifying_2018},
in segments that are no more than approximately 14 days and also contain no gaps longer than 1 day.
Keplerspline uses the Bayesian information criterion to automatically pick the most appropriate breakpoint spacing for each light curve.
The detrended segments are then normalized and joined together to search for additional planets.
Keplerspline estimates the standard deviation $\sigma$ of the light curve from the median absolute deviation.
This $\sigma$ is calculated per sector and is used as the nominal uncertainty for all data points in a given sector
for the subsequent transit search.
We also reject points that lie more than $5\sigma$ above the spline
in order to mitigate the effect of flares and instrumental artifacts
on the transit detection algorithm.

\paragraph{Performing the BLS search}
We use the \cuvarbase{} \citep{2022ascl.soft10030H} package
for BLS search,
similar to QLP's implementation since sector~59
\citep{2023RNAAS...7...28K}.
The \cuvarbase{} package uses the NVIDIA CUDA interface to implement BLS on the GPU.
We search in the period range of 4~hours to 10~days
with the optimal frequency spacing $\delta f \propto f^{2/3}$
and a log-uniform grid of the transit duration fraction
with 20 intervals between $2^{-7}$ and $2^{-2}$.
Since we use a fast implementation of BLS that only outputs the BLS power at each period,
once we obtain the period with the highest BLS power,
we use the \astropy{} package
to obtain the transit parameters
(mid-transit time, transit duration $\transitdur$, and depth $\delta$) at the detected period,
following the methods implemented by \vartools{}
\citep{hartman_vartools_2016}.
These parameters are used in subsequent iterations for signal removal.

\subsection{Vetting transit-like signals}  \label{sec:vetting}

We declared any signal with a BLS signal-to-noise ratio $\snr > 7$ as computed by \astropy{} to be
a threshold-crossing event (TCE),
and all TCEs were vetted to ensure they are likely to be transiting planets.
We applied the following criteria in order to automatically eliminate
TCEs of low data quality or whose parameters are incompatible with being a planet:
\begin{enumerate}
    \item number of points in transit $\ge 20$,
    \item number of transits (each with $\ge 3$ points) $\ge 3$,
    \item $a / R_\star > 3$, where $a$ is the semimajor axis and $R_\star$ is the stellar radius, and
    \item $\transitdur < 3 \, T_\text{dur,circ} $, the transit duration assuming a circular orbit and impact parameter $b = 0$.
\end{enumerate}
Both $a / R_\star$ and $T_\text{dur,circ}$ were calculated from
the stellar masses and radii from TICv8 and the period and transit depth of the TCE.
All TCEs that fulfill these criteria,
as well as TCEs whose stellar parameters are missing from TICv8,
were retained for manual vetting.

We performed manual vetting in two distinct steps.
In the first step, we generated three types of plots for each TCE: the BLS spectrum,
the folded light curve centered on the TCE with binned points and the boxcar transit model,
and undetrended light curves with the detrending spline and the transit signals highlighted.
We used these plots to \added{visually} eliminate TCEs that
\begin{enumerate}
    \item were not transit-like in shape,
    \item showed significant odd--even variations, or
    \item could be attributed to fast varying portions of the light curve
        (e.g. scattered light near the beginning or end of an orbit)
        that the detrending step did not remove.
\end{enumerate}

TCEs that survived the initial manual vetting were subject to additional scrutiny.
We generated additional light curves with enhanced systematics correction from the full frame images by extracting aperture photometry following \citet{Vanderburg2016ApJS} and correcting for spacecraft systematics
using the procedures described by \citet{Vanderburg2019ApJL}.
These additional light curves
were used to generate additional vetting plots,
including a cutout view of the \tess{} FFI
and a simple pixel-level difference image of the median in- and out-of-transit FFIs,
with a customized version of the \giants{} package%
\footnote{The original version is available on GitHub at
\url{https://github.com/nksaunders/giants}.}
\citep{2022AJ....163...53S}.
Since \tess{} has a relatively large pixel size of $21''$,
the photometric aperture could enclose nearby targets that are blended or unresolved
from the true target.
Therefore, we also eliminated TCEs that showed a significant offset in the in- and out-of-transit
difference image
from the expected centroid of the target star.
For the few TCEs that passed all previous vetting steps
but were not known confirmed or candidate planets,
we further used the \textsc{transit-diffImage} package%
\footnote{Available on GitHub at \url{https://github.com/stevepur/transit-diffImage}.}
by S. Bryson
to calculate accurate centroids of the TCE
and the assumed host star
taking account of the pixel response function (PRF).

Finally, we eliminated secondary eclipses by matching them to the period and epoch of the star.
We used the expected depth of the secondary eclipse
as well as the phase offset from the known hot jupiter transits
and manually inspected each light curve to rule out that they could be planets at half or twice the hot jupiter's orbital period.

The TCEs that passed all above vetting steps are the planet detections
listed in \autoref{sec:results}.
These detections are used as the numerator of the occurrence rate in \autoref{sec:occurrence}.

\subsection{Injection--recovery simulation}

In order to evaluate the detection efficiency of the planet search pipeline described in \autoref{sec:bls}
and the vetting procedures described in \autoref{sec:vetting},
we performed injection--recovery tests of simulated transit signals.
For each host star in the hot jupiter sample,
five simulated signals were created via
randomly drawn planetary parameters
from the prior distributions
\begin{subequations}
\begin{align}
    R_\text{p} &\sim \uniformdist(0.5, 10) \, \si{\radius\earth}, \\
    \log P &\sim \uniformdist(\SI{4}{\hour}, \SI{10}{\day}), \\
    b &\sim \uniformdist(\added{0}, 1 + R_\text{p} / R_\star),
\end{align}
\end{subequations}%
\added{where $\uniformdist(a, b)$ is the uniform distribution on the interval $[a, b]$.
To calculate $R_{\text{p}}$ and $b$, stellar masses and radii were either taken from TICv8 when available
or fixed to Solar values otherwise.}%
\footnote{\added{Since the injection--recovery simulation is only used to derive an empirical relationship between theoretical transit S/N and pipeline detection efficiency, the exact choice of stellar parameters here has no bearing on the final occurrence rate as long as the range of S/N is sufficiently sampled.
Rather, the choice of stellar parameters affects the final result through the assumed planet radius prior distribution in Equation~\eqref{eqn:planet_radius_prior}},
where we \emph{did} use the refined stellar parameters from \autoref{sec:params} to derive the imputed transit depth
of each simulated planet.}
All orbits were assumed to be circular,
and we adopted a fixed linear limb darkening law with coefficient $0.3$.
We calculated the simulated signals from these parameters
using the limb-darkened transit light curve model
implemented by the \batman{} Python package,
and the model was supersampled by a factor of
15, 5, and 2 for \tess{} sectors
with FFI exposure times \SI{30}{\minute}, \SI{10}{\minute}, and \SI{200}{\second}, respectively.
Each simulated signal was injected into the corresponding hot jupiter host star's \tess{} light curve,
and we searched the resultant light curve
using the same BLS transit search pipeline as in \autoref{sec:bls}.
As before, we used $\snr > 7$ as the criterion for declaring a BLS signal to be a TCE.

\begin{figure}
    \centering
    \includegraphics[width=\columnwidth]{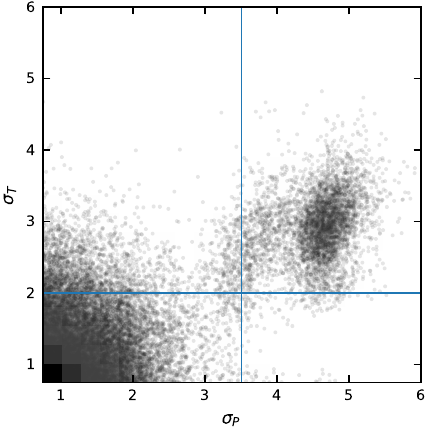}
    \caption{A scatter plot of $\sigma_P$ versus $\sigma_T$ for the detection
    of injected transit signals,
    similar to Figure~2 by \citet{2014AJ....147..119C}.
    The thin blue lines denote the chosen significance levels for ephemeris matching in equation~\eqref{eqn:sigma_pt_levels}
    and divide the plane into four quadrants.
    The cluster of points in the upper right quadrant corresponds
    to true ephemeris matches,
    while the points in the lower left are random false matches.}
    \label{fig:sigma_pt_scatter}
\end{figure}

Rather than manually vetting the TCEs as in \autoref{sec:vetting},
we automatically matched the detected TCEs to the expected period and transit times
using the ephemeris matching method described by \citet{2014AJ....147..119C}.
Using their definition, we set the criteria for ephemeris-matching
to be
\begin{align}  \label{eqn:sigma_pt_levels}
    \sigma_P &> 3.5, &
    \sigma_T &> \added{2.0},
\end{align}
where $P$ is the period and $T$ is the transit time or epoch
and the $\sigma$ values
are converted from the fractional differences in $P$ or $T$
\citep{2014AJ....147..119C}.
\autoref{fig:sigma_pt_scatter},
emulating Figure~2 by \citet{2014AJ....147..119C},
confirms that the chosen significance levels
in Equation~\eqref{eqn:sigma_pt_levels}
capture the mode of true ephemeris matches in the upper right
while excluding the mode of random matches
at low $\sigma_P$ and $\sigma_T$ values in the lower left.
The number of matches as a fraction of injected signals
is the empirical detection efficiency, which we denote $\detecteff$,
and we evaluate its dependence on the transit S/N
in \autoref{sec:detect_eff}.

\subsection{Transit detection efficiency} \label{sec:detect_eff}

\begin{figure}
    \centering
    \includegraphics[width=\columnwidth]{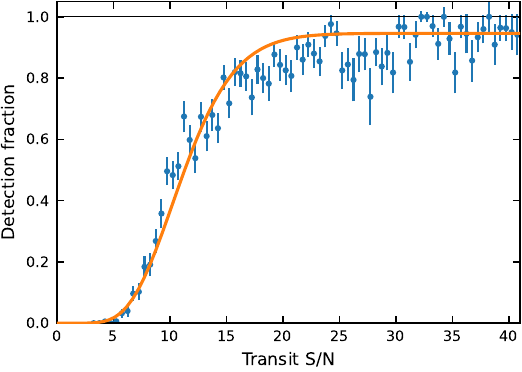}
    \caption{Detection fraction as a function of transit S/N.
    The blue marks are the recovery fraction of injected planets within each S/N bin
    with error bars representing the Agresti--Coull confidence interval of binomial proportion,
    and the orange curve is a scaled gamma distribution CDF.
    The detection fraction asymptotically approaches $\approx 0.95$
    as the transit $\snr \to \infty$.
    }
    \label{fig:snr_detect}
\end{figure}

The injection--recovery simulation
resulted in a map of sets of simulated planetary parameters
to Boolean values indicating success or failure of recovery.
For the convenience of calculating the occurrence rate later in \autoref{sec:occurrence},
we convert the injection--recovery results into an analytical formula 
that expresses the fraction of planets successfully detected by our search
in terms of their transit signal's theoretical S/N.

We calculate a planet's theoretical transit S/N as
\begin{equation}
    \snr = \frac{\delta}{\sqrt{\sigma_\text{sum}^2 P / T_\text{dur} }},
\end{equation}
where $\delta$ is the transit depth,
$P$ is the orbital period,
$T_\text{dur}$ is the transit duration,
and
$\sigma_\text{sum}^2$ is the \enquote{harmonic sum} of $\sigma_i^2$,
the squared uncertainty of the $i$th data point ($i = 1, 2, \ldots, N$)
\begin{equation}
    \sigma_\text{sum}^2 = \bigg( \sum_{i = 1}^N \frac{1}{\sigma_i^2} \bigg)^{-1}.
\end{equation}
In the case that the data points have identical $\sigma_i$, this expression reduces to the familiar
$\sigma_\text{sum}^2 = \sigma_i^2 / N$.

In order to relate the theoretical S/N to the actual detection fraction,
we empirically fitted the cumulative distribution function (CDF) of the gamma distribution
with a shape parameter $\alpha$, scale parameter $\theta$, and maximum detection fraction $C$:
\begin{equation}
    F(x; \alpha, \theta, C) = \frac{C}{\Gamma(\alpha) \, \theta^\alpha}
        \int_0^{x} t^{\alpha - 1} e^{-t / \theta} \dd{t},
\end{equation}
as was used by \citet{2015ApJ...810...95C} for characterizing the transit signal recovery of the \kepler{} pipeline.
To properly account for the statistics of the binned detection fraction,
we used the Agresti--Coull interval \citep{1998AmSta..52..119A}
where the number of trials $n$ and the success fraction $p$
are modified into their estimators
\begin{align}
    \hat{n} &= n + 1,
    &
    \hat{p} &= \frac{1}{\hat{n}} \left( p + \frac{1}{2} \right),
\end{align}
which are used to calculate the interval
\begin{equation}
    \sqrt{\frac{\hat{p} (1 - \hat{p})}{\hat{n}}} .
\end{equation}
We used the unmodified $p$ for each S/N bin
and the Agresti--Coull interval as the uncertainty in each bin.

We ended up with the best-fit parameters
\begin{align}
    \alpha = 9.21719769,
    \theta = 1.24222051,
    C = 0.94609528,
\end{align}
given to 8 decimal places (understood as machine precision and not as significant figures).
This scaled gamma CDF $F(x)$,
illustrated in \autoref{fig:snr_detect},
will be used in \autoref{sec:occurrence}
to convert the theoretical transit S/N of the simulated planets
into the expected pipeline detection efficiency $\detecteff$
by setting $x = \snr$.

\section{Planet Detections} \label{sec:results}

We detected the following inner companions, in ascending order of the orbital period of their respective hot jupiters:
\begin{enumerate}
    \item \toifourteenoheightinner{} \citep{2024ApJ...971L..28K},
    \item \toififtyonefortythreeinner{} \citep{2026AJ....171..359Q},
    \item \toieleventhirtyinner{} \citep{2020ApJ...892L...7H},
    \item \toitwentyfourninetyfourinner{} \citep{2026AJ....171..359Q},
    \item \waspeightyfourinner{} \citep{2023MNRAS.525L..43M}, and
    \item \toitwothousandinner{} \citep{2023MNRAS.524.1113S}.
\end{enumerate}
Of these, \toitwothousandinner{} was discovered by an earlier version of the pipeline described in
\autoref{sec:bls}.
These detections form the basis of the occurrence rate calculation in \autoref{sec:occurrence},
\added{and their system names are highlighted in magenta in \autoref{fig:family_portrait}}.

The following known nearby companions were not among those detected:
\begin{enumerate}
    \item \waspfortyseveninner{} and \waspfortysevenouter{}, the inner and outer companions to \waspfortysevenhj{},
    \item \toifortyfoursixtyeightouter{}, the outer companion to \toifortyfoursixtyeighthj{} \citep{Livesey2026ApJL},
    \item \keplerseventhirtyinner{}, the inner companion to \keplerseventhirtyhj{}, and
    \item \wasponethirtytwoinner{}, the inner companion to \wasponethirtytwohj{}.
\end{enumerate}
The hot jupiters \waspfortysevenhj{} and \toifortyfoursixtyeighthj{} were included in our sample,
but their respective companions
were not detected by the pipeline.
The hot jupiter \keplerseventhirtyhj{} was not a TOI and thus not in the sample.
The hot jupiter \wasponethirtytwohj{} was not in the sample because it lacked uncertainty for its radius
in the TOI table on the ExoFOP website at the time of access.
They will not count as detections for the purpose of the occurrence rate calculation.

\section{Occurrence Rate of Nearby Companions to Hot Jupiters}  \label{sec:occurrence}

Our goal is to estimate the occurrence rate $\eta$ of nearby companions to hot jupiters,
conditioned on the hot jupiter existing.
To this end, we use a Bayesian statistical framework
to calculate the posterior distribution of $\eta$ given our sample of transiting companions to transiting hot jupiters from \tess{}.
We model each hot jupiter observation
as a Bernoulli trial
where the probability of success is $\eta$.
Then, given $\numtrial$ such conditionally independent Bernoulli trials
with $\numobs$ observed successes
and a uniform prior for $\eta$,
the posterior distribution of $\eta$
can be modeled as the beta distribution $\mathrm{Beta}(s + 1, n - s + 1)$,
which is the conjugate prior of the binomial distribution.
The PDF of $\eta$ can be written as
\begin{align}  \label{eqn:beta}
    f(\occrate; s, n)
    &= \frac
        {\occrate^{\numobs} (1 - \occrate)^{\numtrial - \numobs}}
        {\mathrm{B}(\numobs + 1, \numtrial - \numobs + 1)} ,
\end{align}
where $\mathrm{B}$ is the beta function.
We take $s = 6$ as the number of hot jupiters with detected companions (\autoref{sec:results}),
while the number of trials $n$ is the number of hot jupiter systems searched
corrected for observational completeness
and accounting for the possibility of hot jupiter false positives.

Observational completeness concerns two questions
regarding a hypothetical companion to a hot jupiter.
First, given the orbital and planet parameters, does the companion transit?
Second, if the companion transits, how likely is its detection?
In order to evaluate these two questions for each hot jupiter system,
we simulate \num{10000} hypothetical companions with the following procedure.
\begin{enumerate}
    \item The companion's radius, orbital period, and mutual inclination with the hot jupiter are drawn
        from a prior distribution.
    \item The companion's impact parameter is calculated from the mutual inclination
        and the inclination of the hot jupiter.
    \item If the impact parameter indicates that the companion does not transit, the detection efficiency $\detecteff$ is set to zero,
        and we move on to the next simulated planet.
        Otherwise, we calculate the expected transit S/N and convert it into a predicted
        efficiency as 
        $\detecteff = F(\snr)$,
        with $F$ being the empirically derived gamma CDF in \autoref{sec:detect_eff}.
\end{enumerate}
Finally, we sum up the average detection efficiency of the simulated companions in each hot jupiter system
to arrive at an effective sample size (ESS) of the hot jupiters searched,
which we take to be the number of trials $n$.
The rest of this section describes each step in detail.

\subsection{Prior distribution of nearby companions}  \label{sec:occ_priors}

We draw the companion's radius $R_\text{p}$, orbital period $P$, and mutual inclination $\psi$
from the following distributions:
\begin{subequations}
\begin{align}
    R_\text{p} &\sim \uniformdist(1, 4)\,\si{\radius\earth} , \\  \label{eqn:planet_radius_prior}
    P &\sim \uniformdist(0.25, 10)\,\si{\day} , \\
    \psi &\sim \mathrm{Rayleigh}(\ang{1.8;;}) ,
\end{align}
\end{subequations}
where $\uniformdist(a, b)$ is the uniform distribution on the interval $[a, b]$
and $\mathrm{Rayleigh}(\sigma)$ is the Rayleigh distribution with the scale parameter $\sigma$,
subject to the condition
\begin{equation}
    a / R_\star > 2 .
\end{equation}
For ease of comparison, we follow
\citet{2016ApJ...825...98H}
in setting $\sigma$ to \ang{1.8;;} (\added{converted to} radians),
a value which was in turn based on an estimate of \ang{1.0}--\ang{2.2}
for $\psi$ in \kepler{} multi-transiting systems
by \citet{Fabrycky2014ApJ}.
We will explore how the inferred occurrence rate changes
with the choice of prior distributions for $\psi$ in \autoref{sec:mutual_inc}.

\subsection{Mutual inclination and impact parameter}

We calculate the impact parameter of the simulated companion from its $\psi$
in order to find its transit duration, on which the transit S/N depends.
\added{We sample the companion's inclination $i_2$ through the relation}
\begin{equation}
    \cos i_2 = \cos i_1 \cos \psi + \cos B \sin i_1 \sin \psi ,
\end{equation}
where $i_1$ is the inclination of the hot jupiter derived from its impact parameter
(see \autoref{sec:params})
\added{and $B$ is an angle randomly drawn from $\uniformdist(0, \ang{360})$}.%
\footnote{\added{If we set the orbital normal vector of the hot jupiter as the $z$-axis, then $\psi$ and $B$ are respectively the polar and azimuthal angles of the companion's orbital normal vector.}}
The companion's impact parameter $b_2$ is then calculated from $\cos i_2$
assuming a circular orbit.

If $b_2 > 1 + R_\text{p} / R_\star$, the planet is deemed not transiting, and its detection efficiency is set to zero.
Otherwise, we proceed to calculate its transit S/N and detection efficiency as in \autoref{sec:detect_eff}
using the stellar radii from \autoref{sec:params}.

\begin{figure*}
    \centering
    \includegraphics[width=0.8\textwidth]{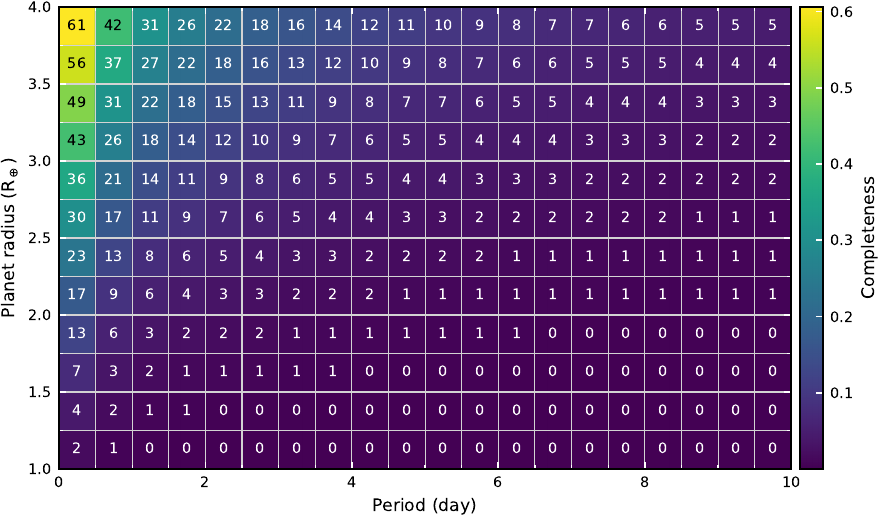}
    \caption{Overall observational completeness of simulated nearby companions to hot jupiters
    in bins of the orbital periods and planet radii of the companions (\autoref{sec:completeness}), expressed as percentages.
    The observational completeness takes account of both transit probability
    and detection efficiency.
    }
    \label{fig:detect_eff}
\end{figure*}

\subsection{Observational completeness and effective sample size}  \label{sec:completeness}

To calculate an effective sample size (ESS),
we sum up the average detection efficiency $\detecteff$ of the simulated companions in each hot jupiter system.
In mathematical notation,
the total ESS is
\begin{equation}
    \mathrm{ESS} = \sum_{j = 1}^{N_\text{HJ}} \frac{1}{N_\text{sim}} \sum_{i = 1}^{N_\text{sim}} \detecteff_{ij} ,
\end{equation}
where $\detecteff_{ij}$ is the detection efficiency of
the \added{$i$th simulated companion of the $j$th hot jupiter},
$N_\text{sim} = \num{10000}$ is the number of simulated companions per hot jupiter,
and $N_\text{HJ}$ is the number of hot jupiter systems in our sample.
\autoref{fig:detect_eff} shows the average $\detecteff_{ij}$
in bins of planet period and radius
expressed as percentages.
Note that $\detecteff_{ij} \equiv 0$ for non-transiting simulated companions,
so the ESS is a measure of observational completeness taking account of both
transit probability and pipeline detection efficiency.

Some of the hot jupiter candidates in our sample from \tess{} surveys
may still be false positives (e.g. background eclipsing binaries,
hierarchical triples, blended foreground binaries).
We multiply the above $\mathrm{ESS} = 92.50$ by $(1 - \mathrm{FPR}_\text{HJ})$ from \autoref{sec:fpr}
to obtain the final $\numtrial = 86.95$,
which is substituted into equation~\eqref{eqn:beta}
to obtain the posterior distribution PDF $f(\eta)$.

\begin{figure}
    \centering
    \includegraphics[width=\columnwidth]{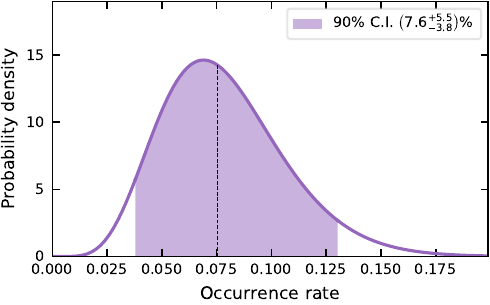}
    \caption{Posterior distribution of the occurrence rate of nearby companions
    ($\SI{0.25}{\day} \le P < \SI{10}{\day}$, $1 \le R_\text{p} / \si{\radius\earth} < 4$)
    to hot jupiters.
    The 90\% credible interval about the median is \occRateHjAll{},
    with the median indicated by a solid vertical line and the interval shaded.
    }
    \label{fig:occ}
\end{figure}

\subsection{The occurrence rate} \label{sec:occ}

The PDF of the posterior distribution $f(\eta)$
of the occurrence rate $\eta$ of nearby companions to hot jupiters
is shown in \autoref{fig:occ}.
We quote the median and 90\% credible interval
of the occurrence rate as
\occRateHjAll{}.

\subsection{Dependence on mutual inclination}  \label{sec:mutual_inc}

In order to test the effect of orbital alignment between the hot jupiter and its nearby companion
on the occurrence rate,
we replace the Rayleigh distribution used in \autoref{sec:occ_priors}
with a series of 3D von Mises--Fisher (VMF) distributions of varying parameters.
The 3D VMF distribution
is the analog to the 2D normal distribution on the sphere,
with a PDF over the unit vector $\hat{\vb{h}}$
\begin{equation}
    f(\hat{\vb{h}}; \hat{\vb{n}}, \kappa) = \frac{\kappa}{4\pi \sinh{\kappa}} \exp(\kappa \hat{\vb{n}} \cdot \hat{\vb{h}}) ,
\end{equation}
with $\hat{\vb{n}}$ being a unit vector representing the reference direction
and $\kappa$ a scale parameter characterizing the spread of the distribution.
As such, if we take $\hat{\vb{n}}$ and $\hat{\vb{h}}$
to be the orbit normal vector of the hot jupiter and its nearby companion, respectively,
then the VMF distribution is a good candidate for parameterizing
their degree of alignment via the scale parameter $\kappa$.
If we marginalize this distribution over \added{the azimuthal angle about $\hat{\vb{n}}$, leaving only the mutual inclination $\psi$}
(\added{with} $\cos\psi = \hat{\vb{n}} \cdot \hat{\vb{h}}$),
then we obtain the Fisher distribution with a PDF of
\begin{equation}
    f(\psi; \kappa) = \frac{\kappa}{2 \sinh{\kappa}} e^{\kappa \cos{\psi}} \sin{\psi} ,
\end{equation}
which is the form often used in theoretical treatments of orbital alignment in multiplanet systems
(e.g. \citealt{2017AJ....154..230B,2009ApJ...696.1230F}).
The Fisher distribution with scale parameter $\kappa$ reduces to $\mathrm{Rayleigh}(1/\sqrt{\kappa})$
for large $\kappa$ and small $\psi$
and to an isotropic distribution
(i.e. uniform in $\cos\psi$)
as $\kappa \to 0$,
representing the cases of well-alignment versus maximal misalignment, respectively.

We find that as $\kappa \to 0$, we have $\numtrial \approx 17$,
and as $\kappa \to \infty$, $\numtrial \approx 89$.
This translates to an occurrence rate of
\occRateHjIsotropic{}
if the companion is isotropically distributed and
\occRateHjAligned{}
if the companion is aligned.

\begin{figure*}
    \centering
    \includegraphics[width=\textwidth]{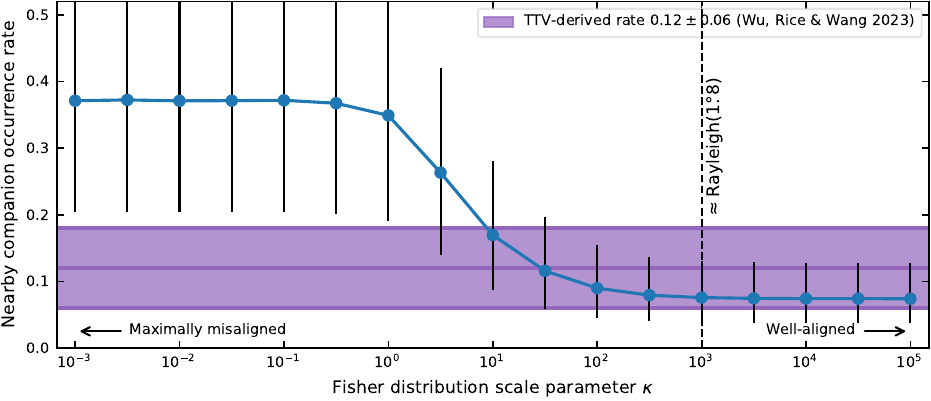}
    \caption{Occurrence rate of nearby companions
    ($\SI{0.25}{\day} \le P < \SI{10}{\day}$, $1 \le R_\text{p} / \si{\radius\earth} < 4$)
    to hot jupiters as a function of their orbital alignment,
    parameterized as the scale parameter $\kappa$ of a Fisher distribution
    prior on their mutual inclination.
    The posterior of each occurrence rate is a beta distribution,
    and we take the median and 90\% interval as the marks and error bars shown.
    The vertical line ($\kappa \approx 10^3$)
    corresponds to the $\mathrm{Rayleigh}(\ang{1.8;;})$ distribution,
    which is used as the mutual inclination prior to derive our main occurrence rate result of
    \occRateHjAll{}.
    For comparison, we shade the $(12 \pm 6) \%$ credible interval of
    a similar occurrence rate derived from transit-timing variations of hot jupiters from the \kepler{} mission
    by \citet{2023AJ....165..171W}.
    While their occurrence rate is not directly comparable to ours,
    reconciling the two different detection methods
    appears to rule out the case of maximal misalignment between hot jupiters and their nearby companions.
    }
    \label{fig:mutual_inc}
\end{figure*}

\section{Discussion}
\label{sec:discussion}

In this work, we measure the occurrence rate of \added{small (radius 1--\SI{4}{\radius\earth}) nearby (orbital period 0.25--\SI{10}{\day})} companions to hot jupiters to be \occRateHjAll. In agreement with previous estimates of the occurrence rates of nearby companions \citep{2016ApJ...825...98H, Hord2021AJ}, we find that hot jupiters tend to be solitary. Historically, this fact has been taken as evidence that hot jupiter formation is dynamically violent and tends to destabilize nearby planetary companions. In this section, we revisit these conclusions in light of our new results.

\subsection{Comparison to inner companions of small short-period planets}  \label{sec:compare_small_planets}

Do hot jupiters have fewer nearby companions than the typical small, short-period transiting planet,
such as the ones observed by \kepler{}?
Answering this question puts the loneliness of hot jupiters
into the broader context of planetary system architectures.
Using data from \kepler{} DR25 \citep{2018ApJS..235...38T},
we calculated the probability of a planetary system
hosting small transiting short-period planets
also hosting another transiting small planet nearby ($P < \SI{10}{\day}$).
Specifically, we selected small planets with
\begin{enumerate}
    \item \added{host star effective temperatures} $T_{\text{eff},\star} > \SI{4500}{\kelvin}$, 
    \item radii $R_\text{p} < \SI{4}{\radius\earth}$ (assuming \gaia{} DR2 stellar parameters),
    \item periods $\SI{0.25}{\day} \le P < \SI{10}{\day}$, and
    \item impact parameters $b < 0.9$.
\end{enumerate}
We set an additional stellar effective temperature criterion to make sure the overall stellar sample is similar to that of hot jupiter hosts. 
We found this probability is $17.5\%$ for the small planets even without correcting for completeness, 
much higher than the intrinsic occurrence rate for nearby companions to hot jupiters.

We further estimated a preliminary intrinsic occurrence rate of these inner companions
using the same prior distribution of planetary parameters for the companions of hot jupiters described in \autoref{sec:occ_priors}. 
Combined with the detection completeness functions provided by the \kepler{} team
\citep{2015ApJ...809....8B,2015ApJ...810...95C},
we find that, on average, the probability of a system hosting a short-period small planet also hosting another small planet within 10 days is $(77\pm5)\%$.
We caution that these results are highly sensitive to the choice of prior distributions
for the companion's period and radius, which is not taken into account in the uncertainties,
although it would not change the conclusion
that the probability of having a companion is significantly higher for small short-period planets compared to hot jupiters.

\subsection{Comparing to occurrence rate obtained from transit-timing variation}

Another important clue we can glean from our analysis is a constraint on the mutual inclination distribution for hot jupiters and their close companions. The mutual inclination between a hot jupiter and its nearby companion tells us a great deal
about the formation and migration history of these systems.
If the hot jupiter and its nearby companion are well-aligned,
then it is evidence that the system had a relatively quiescent dynamical history.
While we can directly constrain the sky-projected orbital orientation of the hot jupiter
through Rossiter--McLaughlin measurements
\citep{1924ApJ....60...15R,1924ApJ....60...22M},
it is challenging to achieve the S/N
required for the small companion within its transit duration.
Thus, we are unlikely to directly constrain the mutual inclination between the hot jupiter and its nearby companion in the near future.
The assumed mutual inclination distribution directly affects the occurrence rate we measure
by altering the observational completeness of nearby companions,
because they must transit in order to be detected in our search.
Having explored
how the degree of alignment between hot jupiters and their nearby companions
affects the calculated occurrence rate in \autoref{sec:mutual_inc},
we now compare our result to occurrence rates derived from
transit-timing variations (TTVs) of hot jupiters,
which can detect non-transiting companions
by their gravitational influence on a hot jupiter's orbit.

After a comprehensive search for TTVs among all hot jupiters observed by \kepler{},
\citet*{2023AJ....165..171W} found two TTV signals among 50 hot jupiters in their sample,
implying an occurrence rate of 
$(12 \pm 6)\%$
for nearby companions to hot jupiters,
where \enquote{nearby} is defined as having a period ratio $\approx 1.5$--4.
Their result is highlighted in \autoref{fig:mutual_inc}.
Although their result is not directly comparable to ours
due to differences in the assumption of priors for the underlying
distribution of nearby companions
and the different detection characteristics of TTV- and transit-based methods,
reconciling the rates derived from the two methods
appears to be incompatible with the case of maximal misalignment between hot jupiters and their nearby companions
(i.e. $\kappa \ll 1$).
This would rule out unusual formation scenarios that result in the hot jupiter and its companion
being in nearly perpendicular orbits,
as would be the case suggested by \citet{Batygin2016ApJ}.
A more careful comparison between the TTV- and transit-based methods
would require a joint search for TTVs and transits among
a complete sample of well-characterized \kepler{} and \tess{} hot jupiters.
As a first step towards this process,
we note that \citet{2024ApJS..275...32Z} have performed a TTV search among 260 hot jupiters from \tess{},
constraining the upper mass limit for potential companions to be several $\si{\mass\jupiter}$.
A future work could combine and extend the results from \citet{2023AJ....165..171W},
\citet{2024ApJS..275...32Z}, and this work
and invert the statistical framework used to derive occurrence rates
to instead constrain the underlying distribution
of mutual inclinations in systems hosting hot jupiters with nearby companions.

\subsection{Do hot jupiter inner companions have a small preferential misalignment?} \label{sec:grazing}

We can also learn about the mutual inclination distribution of the hot jupiters and their close neighbors by studying the impact parameter distributions of the systems in our sample. We note that among the nine hot jupiters with \emph{inner} companions,
four of them
(TOI-1408, TOI-5143, TOI-2494, and TOI-1130)
have grazing transits \added{(these systems are marked with an asterisk in \autoref{fig:family_portrait})}.
Whether a single transiting planet is grazing depends only on
the geometry of an observer viewing a randomly oriented orbital plane.
In a system of two transiting planets, however,
the probability of either of the planets grazing
also depends on their mutual inclination.
If the two planets are perfectly aligned ($\psi = 0$),
then the probability of finding a grazing planet is independent of the existence of the other transiting planet.
But if the two planets are slightly misaligned,
then the probability of finding one of the planets grazing
\added{given the presence of another transiting planet in the same system}
is boosted \added{from what one might expect by random alignment}.
Here, we investigate whether finding four out of nine grazing hot jupiters is more than what chance alignment can explain.

In order to calculate the grazing probability,
we took the planet-to-stellar radius ratio of the $i$th hot jupiter $k_i = R_{\text{p},i} / R_{\star,i}$
and calculated the grazing probability $p_i$ as
\begin{equation}
    p_i = \frac{2 k_i}{1 + k_i} .
\end{equation}
Then, we averaged the grazing probability and used the average
$p = ( \sum_{i = 1}^{9} p_i ) / 9 $ as the parameter of a binomial distribution with the
PDF
\begin{equation}
    f_{\mathrm{binom}}(k; n, p) = \binom{n}{k} p^k (1 - p)^{n - k} ,
\end{equation}
where $n = 9$.
Since the four planets are grazing, their radii are not known precisely,
so we assume $R_\text{p} = \SI{1}{\radius\jupiter}$.
This gives an average grazing probability $p = 0.19$,
which implies that the probability of having 4 or more grazing planets
\added{due to random viewing geometry}
is
\begin{equation}
    \Pr(\ge 4 \text{ grazing}) = \sum_{k=4}^9 f_{\text{binom}}(k; 9, 0.19) \approx 0.075 .
\end{equation}
This tantalizing statistical hint merits more investigation as we detect more systems in the future.
If hot jupiters with transiting inner companions are more likely to be grazing than we would expect
from the random orientation of the viewing geometry alone,
as would be the case for solitary hot jupiters without nearby companions,
then there may be a systematic small misalignment between these hot jupiters and their inner companions.
This misalignment could be the remnant dynamic signature
of past migration.

\subsection{Hot jupiters with inner companions also have cold outer companions}

We can also investigate the systems with close companions and study their architectures beyond short-period transiting systems. Although transits are an effective way of finding additional planets in hot jupiter systems,
they may not give us the full picture of the architecture of these systems
as transit probability falls off for longer orbital periods.
Indeed, using up to \SI{30}{\year} of RV measurements from the California Legacy Survey,
\citet{2023ApJ...956L..29Z}
found that virtually all hot jupiters have at least one more \enquote{cold} companion
at much greater orbital periods.
Since the very presence of these cold companions likely caused the hot jupiters
to migrate inward after disk dispersal,
it would be instructive to look for cold companions in hot jupiter systems
with nearby companions
to see whether they differ from the hot jupiter population at large.
Here, we assemble evidence from the literature for cold outer companions.
\begin{enumerate}
    \item \emph{WASP-47}.
        \citet{2016A&A...586A..93N} found a cold companion \waspfortysevencold{}
        with orbital period $P = \SI{588.8\pm2.0}{\day}$
        and minimum mass $M_\text{p} \sin i = \SI{398.9 \pm 9.1}{\mass\earth}$
        \citep{2022AJ....163..197B}.
        Additionally, \waspfortysevenouter{} is a nearby outer companion
        \citep{2015ApJ...812L..18B}.
    \item \emph{WASP-132}.
        Planet d is detected by RV with $P = \SI[parse-numbers=false]{(1811.8^{+42.6}_{-44.4})}{\day}$,
        $M_\text{p} \sin i = \SI[parse-numbers=false]{(5.29^{+0.48}_{-0.46})}{\mass\jupiter}$ \citep{2025A&A...693A.144G}.
        There is an additional linear trend in the RV
        (with a baseline of $\approx \SI{9}{\year}$ implying $P \gtrsim \SI{18}{\year}$)
        that may be a brown dwarf,
        in line with evidence for excess astrometric uncertainty
        in \gaia{} DR2 and DR3.
    \item \emph{TOI-1130}.
        There is an additional linear trend
        implying $P \gtrsim \SI{160}{\day}$ \citep{2024A&A...689A..52B}.
    \item \emph{TOI-1408}.
        There is an additional nonlinear trend
        implying $P \approx \SI{2530}{\day}$, $M \approx \SI{14.6 \pm 0.3}{\mass\jupiter}$ \citep{2024ApJ...971L..28K}.
\end{enumerate}
In addition, \toitwothousand{} showed
planet-sized RV signals at $\approx\SI{17}{\day}$ and $\approx\SI{90}{\day}$
not explained by spectroscopic stellar activity indicators
or photometric variations
\citep{2023MNRAS.524.1113S},
although the RV baseline was too short
to draw any conclusions at the time.
Since then, \added{the High Accuracy Radial velocity Planet Searcher (HARPS) spectrograph
at the ESO La Silla \SI{3.6}{\meter} telescope}
has made 39 additional RV measurements
from 2023 December to 2024 March,%
\footnote{ESO program 112.25WB, PI: Chelsea X. Huang.}
which showed an overall offset from the RVs three years prior in 2021
at the \SI{20}{\meter\per\second} level,
although it is too early to say whether this offset
is due to long-term stellar activity
or an additional long-period planet.

Given that these systems are newly characterized
and lack long-term RV monitoring,
it is quite possible that there are cold outer companions
in the remaining systems
whose signals are hiding in plain sight,
which would imply that
hot jupiters with nearby companions
seem to also have cold companions
at a rate comparable to that for other hot jupiters in general estimated by
\citet{2023ApJ...956L..29Z}.
The presence of outer companions is believed to be a viable way to kickstart high-eccentricity migration (HEM),
but if hot jupiters with nearby companions, which presumably did not undergo HEM, also have a similar (or even greater) prevalence of outer companions,
then the link between outer companions and HEM would be more complicated than it currently appears;
\added{namely, the mere presence of a cold companion is not sufficient to trigger HEM in producing a hot jupiter}.
Continued long-term RV monitoring of these systems can potentially reveal this tension
\added{by comparing the mass and eccentricity of these suspected cold companions
to the distribution of other hot jupiters' cold companions}.

\subsection{Implications on formation theories}

In summary, we find that about \occRateHjAll\ of all hot jupiters have closely orbiting companion planets, that the mutual inclination of these planets is small, but nonzero and probably larger than the mutual inclination of small planet systems, and that the frequency of long-period outer companions seems similar to that of hot jupiters without close companion planets. We draw the following conclusions about the formation of hot jupiters. 

\begin{enumerate}
    \item At least \occRateHjAll\ of hot jupiters (i.e. the ones with nearby companions) must not form via high-eccentricity migration, which would almost certainly disrupt any nearby planetary companions. Moreover, the likely low mutual inclinations between hot jupiters and their close companions also suggest relatively quiescent formation scenarios. 
    \item Moreover, since \occRateHjAll\ is a strict lower limit on the fraction of systems that formed via dynamically quiescent mechanisms, the true fraction of hot jupiters that form in this way could be larger. For example, simulations of hot jupiters with nearby companions that formed via disk migration indicate that only about 1/3 of such close companions survive for long timescales \citep{Wu2023AJ}. This means that our result that \occRateHjAll\ of hot jupiters have close companions is consistent with a larger fraction ($\approx 20\%$) of hot jupiter systems forming via disk migration. 

\end{enumerate}

Going forward, more detailed measurements and studies of the architectures of these systems, including the occurrence rate of specifically ultra-short-period planets \citep{Wang2025ApJL}, stellar obliquities and mutual inclinations \citep{He2024MNRAS}, and long-term RV surveys to measure the frequency of long-period massive planets \citep{2023ApJ...956L..29Z}, will make it possible to estimate the relative importance of the different hot jupiter formation channels.

\section{Conclusions}  \label{sec:conclusions}

Using five years of \tess{} data,
we have performed the most comprehensive search
for nearby transiting companion planets to hot jupiters
to date.
After accounting for observational completeness and pipeline detection efficiency,
we arrived at an intrinsic occurrence rate of
\occRateHjAll{}
for nearby companions to hot jupiters.
This result is highly sensitive to the \added{assumed} mutual inclination distribution
between the hot jupiters and their companions,
and we instead have
\occRateHjIsotropic{}
for the case of maximal misalignment and
\occRateHjAligned{}
for the case of perfect alignment.
Comparing our rate based on transit detections
to that based on TTV detections
appears to disfavor the case of maximal misalignment,
\added{but there may nevertheless be a preference for small misalignment
as there is a hint that more hot jupiters with nearby companions are grazing than can be explained by random viewing geometry.
A review of RVs in the literature reveals that about half of the known hot jupiters with nearby companions
either have or are suspected to have cold companions,
raising questions about the role cold companions play in triggering the formation of hot jupiters via HEM.}
A future comprehensive search for nearby companions
with transits, TTVs, and RVs
may elucidate the underlying mutual inclination distribution.
The picture is emerging that these nearby companions to hot jupiters,
while a relatively uncommon outcome of hot jupiter formation,
provide a key constraint on the \added{relative frequency} of each formation channel.

\begin{acknowledgments}

\added{We thank Joshua N. Winn for suggestions that led to our implementation of \autoref{sec:compare_small_planets}.}

Funding for the TESS mission is provided by NASA's Science Mission Directorate.
We acknowledge the use of public TESS data from pipelines at the TESS Science Office and at the TESS Science Processing Operations Center.
This research has made use of the Exoplanet Follow-up Observation Program website \citep[ExoFOP;][]{exofop5}, which is operated by the California Institute of Technology, under contract with the National Aeronautics and Space Administration under the Exoplanet Exploration Program.
This paper includes data collected by the TESS mission that are publicly available from the Mikulski Archive for Space Telescopes (MAST).

This work makes use of observations from the LCOGT network. Part of the LCOGT telescope time was granted by NOIRLab through the Mid-Scale Innovations Program (MSIP). MSIP is funded by NSF. This paper is based on observations made with the Las Cumbres Observatory’s education network telescopes that were upgraded through generous support from the Gordon and Betty Moore Foundation.

This work is partly supported by JSPS KAKENHI Grant Numbers JP24H00017, JP24K00689, JP24H00248, JP24K17082, JP24K17083, JSPS Grant-in-Aid for JSPS Fellows Grant Numbers JP24KJ0241, JP25KJ0091, and JSPS Bilateral Program Number JPJSBP120249910.
This paper is based on observations made with the MuSCAT instruments, developed by the Astrobiology Center (ABC) in Japan, the University of Tokyo, and Las Cumbres Observatory (LCOGT).
This article is based on observations made with the MuSCAT2 instrument, developed by ABC, at Telescopio Carlos Sánchez operated on the island of Tenerife by the IAC in the Spanish Observatorio del Teide.
MuSCAT3 was developed with financial support by JSPS KAKENHI (JP18H05439) and JST PRESTO (JPMJPR1775), and is located at the Faulkes Telescope North on Maui, HI (USA), operated by LCOGT. MuSCAT4 was developed with financial support provided by the Heising-Simons Foundation (grant 2022-3611), JST grant number JPMJCR1761, and the ABC in Japan, and is located at the Faulkes Telescope South at Siding Spring Observatory (Australia), operated by LCOGT.

Funding for KB was provided by the European Union (ERC AdG SUBSTELLAR, GA 101054354).

AB
acknowledges the support of M.V. Lomonosov Moscow State University Program of Development.

ZLD would like to thank the generous support of the MIT Presidential Fellowship, the MIT Collamore--Rogers Fellowship and to acknowledge that this material is based upon work supported by the National Science Foundation Graduate Research Fellowship under Grant No. 1745302.

KAC acknowledges support from the TESS mission via subaward s3449 from MIT.

TG acknowledges that ASTEP benefited from the support of the French and Italian polar agencies IPEV and PNRA in the framework of the Concordia station program and from OCA, INSU, Idex UCAJEDI (ANR-15-IDEX-01), ESA through the Science Faculty of the European Space Research and Technology Centre (ESTEC), the European Research Council (ERC) under the European Union's Horizon 2020 research and innovation program (grant agreement nº 803193/BEBOP), and from the Science and Technology Facilities Council (STFC; grant nº ST/S00193X/1, ST/W002582/1, and ST/Y001710/1).

FM acknowledges the financial support from the Agencia Estatal de Investigaci\'{o}n del Ministerio de Ciencia, Innovaci\'{o}n y Universidades (MCIU/AEI) through grant PID2023-152906NA-I00.

EP acknowledges financial support from the Agencia Estatal de Investigaci\'on of the Ministerio de Ciencia e Innovaci\'on MCIN/AEI/10.13039/501100011033 and the ERDF “A way of making Europe” through project PID2021-125627OB-C32, and from the Centre of Excellence “Severo Ochoa” award to the Instituto de Astrofisica de Canarias.

\end{acknowledgments}

\software{%
    Astropy \citep{astr13,astr18,2022ApJ...935..167A},
    MIT Quick Look Pipeline (QLP) \citep{mastqlp,2020RNAAS...4..204H,2020RNAAS...4..206H,2021RNAAS...5..234K,2022RNAAS...6..236K},
    giants \citep{2022AJ....163...53S},
    tess-diffImage,
    Pandas,
}

\facilities{\gaia{}, \tess{}}

\bibliographystyle{aasjournalv7}
\bibliography{hjcompanion}

\global\suppressAffiliationsfalse
\allauthors

\end{document}